\documentclass[adp,fleqn]{w-art}
\usepackage{times}
\usepackage{w-thm}
\usepackage{graphicx}
\usepackage{cite}
\usepackage{amsmath,amssymb}

\newcommand{\void}[1]{}
\renewcommand{\d}{\mathrm{d}}
\newcommand{\e}{\mathrm{e}}
\renewcommand{\i}{\mathrm{i}}
\newcommand{\T}{T}
\newcommand{\tr}{\mathop{\mathrm{tr}}\nolimits}
\renewcommand{\Re}{\mathop{\mathrm{Re}}\nolimits}
\renewcommand{\Im}{\mathop{\mathrm{Im}}\nolimits}
\renewcommand{\L}{\mathrm{L}}
\newcommand{\R}{\mathrm{R}}

\begin{document}
\DOIsuffix{theDOIsuffix}
\Volume{xx}
\Issue{x}
\Copyrightissue{xx}
\Month{10}
\Year{2007}
\pagespan{1}{}
\Receiveddate{\today}

\title{Shot noise in non-adiabatically driven nanoscale conductors}

\author[F. J. Kaiser]{Franz J. Kaiser}
\author[S. Kohler]{Sigmund Kohler\footnote{Corresponding
     author: S.~Kohler, e-mail: {\sf sigmund.kohler@physik.uni-augsburg.de},
     Phone: +49\,821\,598\,3316,
     Fax: +49\,821\,598\,3222}}
\address{Institut f\"ur Physik, Universit\"at Augsburg,
Universit\"atsstra{\ss}e 1, 86135 Augsburg, Germany}

\begin{abstract}
We investigate the noise properties of pump currents through molecular
wires and coupled quantum dots.  As a model we employ a two level
system that is connected to electron reservoirs and is
non-adiabatically driven.  Concerning the electron-electron
interaction, we focus on two limits: non-interacting electrons
and strong Coulomb repulsion.  While the former case is treated within
a Floquet scattering formalism, we derive for the
latter case a master equation formalism for the computation of the
current and the zero-frequency noise.
For a pump operated close to internal resonances, the differences
between the non-interacting and the strongly interacting limit turn
out to be surprisingly small.
\end{abstract}

\maketitle

\section{Introduction}

Recent experiments with coherently coupled quantum dots
\cite{Blick1996a, vanderWiel1999a, vanderWiel2003a, Khrapai2006a} and
molecular wires \cite{Cui2001a, Reichert2002a} deal with the transport
properties of small systems with a discrete level structure.  These
experimental achievements generated new theoretical interest in the transport
properties of nanoscale systems \cite{Nitzan2001a, Hanggi2002article}.  One
particular field of interest is the interplay of transport and
electronic excitations by an oscillating gate voltage, a microwave
field, or an infrared laser, respectively.  Such excitations bear
intriguing phenomena like photon-assisted tunnelling \cite{Tien1963a,
Inarrea1994a, Blick1995a, Stafford1996a, Brune1997a, Hazelzet2001a,
vanderWiel2003a, Platero2004a, Kohler2005a} and the suppression of
both the dc current \cite{Lehmann2003a, Kleinekathofer2006a} and the
zero-frequency noise \cite{Camalet2003a, Kohler2004a}.

A further intriguing phenomenon in this context is electron pumping
induced by a cyclic change of the parameters in the absence of any
external bias voltage \cite{Kouwenhoven1991a, Pothier1992a,
Switkes1999a}.  For adiabatically slow driving, the transfered charge
per cycle is determined by the area enclosed in parameter space during
the cyclic evolution \cite{Brouwer1998a, Altshuler1999a}.  This
implies that the resulting current is proportional to the driving
frequency and, thus, suggests that non-adiabatic electron pumping is
more effective.
For practical applications, it is also desirable that the pump current
flows at a sufficiently low noise level.  It has been found that
adiabatic pumps can be practically noiseless \cite{Avron2001a}.  This
happens, however, on the expense of having only a small or even
vanishing current \cite{Polianski2002a}.  Outside the adiabatic
regime, when the driving frequency is close to the internal resonances
of the conductor, the current assumes much larger values while its
noise nevertheless is clearly sub-Poissonian~\cite{Strass2005b}.
Since this prediction of an optimal working point has been made for
non-interacting electrons, the question on the influence of Coulomb
repulsion arises.

An intuitive description of the electron transport through
mesoscopic systems is provided by the Landauer scattering formula
\cite{Landauer1957a, Blanter2000a} and its various generalisations.
In this formalism, both the average current \cite{Datta1995a} and the
transport noise characteristics \cite{Blanter2000a, Chen2003a} can be
expressed in terms of the quantum transmission probabilities of the
respective scattering channels.  If one heuristically postulates that
the current obeys a scattering formula, one should worry whether this
complies with the Pauli principle or if it has to be ensured by
introducing ``blocking factors'' \cite{Datta1995a}.  For
static conductors the current being
the experimentally relevant quantity, is independent of these blocking
factors, which renders this question rather academic.  This is no
longer the case when the scattering potential is time-dependent.  Then a
scattered electron can absorb or emit energy quanta from the driving
field, which opens inelastic transport channels \cite{Wagner1995a,
Li1999a, Henseler2000a}.  So blocking factors indeed
can have a net effect on the current, and it has been suggested to test
the demand for them experimentally with driven conductors
\cite{Datta1992a, Wagner2000a}.
In order to avoid such conflicts, one should start from a
many-particle description.  In this spirit, within a Green function
approach, a formal solution for the current through a time-dependent
conductor has been presented, e.g., in Refs.~\cite{Datta1992a} and
\cite{Jauho1994a} without taking advantage of the full Floquet theory
for the wire.  Nevertheless in some special cases like, e.g., for
conductors consisting of only a single level \cite{Wingreen1993a,
Aguado1996a} or for the scattering by a piecewise constant potential
\cite{Wagner1999a}, an explicit solution becomes
feasible.  A complete Floquet theory provides in addition to a current
formula a prescription for the computation of the Green function
\cite{Camalet2004a, Kohler2005a}.

The spectral density of the current fluctuations has been derived for the
low-frequency ac conductance \cite{Pretre1996a, Pedersen1998a} and the
scattering by a slowly time-dependent potential \cite{Lesovik1994a}.  For
arbitrary driving frequencies, the noise has been characterised by its
zero-frequency component \cite{Camalet2003a}.  A remarkable feature of the
current noise in the presence of time-dependent fields is its dependence on
the phase of the transmission amplitudes \cite{Lesovik1994a,
Camalet2003a}.  By clear contrast, both the current in the driven
case \cite{Camalet2003a} and the noise in the static case
\cite{Blanter2000a} depend solely on transmission probabilities.

When electron-electron interactions beyond the mean-field level become
relevant, the direct application of a Landauer-like theory is no longer
possible and one has to resort to other methods like, e.g., a master
equation description for the reduced density operator of the wire
\cite{Gurvitz1996a, Petrov2001a, Petrov2002a, Lehmann2002a}.  For
time-dependent conductors, this enables a rather efficient treatment
of transport problems after decomposing the wire density operator into
a Floquet basis.  Then it is possible to study relatively large driven
conductors \cite{Kohler2005a} and to include also electron-electron
interactions \cite{Bruder1994a, Stoof1996a} and electron-phonon
interactions \cite{Lehmann2004a}.
For the computation of the current fluctuations, one can employ a
generalised master equation that resolves the number of the
transported electrons.  This degree of freedom is traced out after
introducing a counting variable \cite{Elattari2002a}.  For various
static transport problems, this approach has been followed by several
groups \cite{Bagrets2003a, Kiesslich2003a, Novotny2004a, Flindt2004a,
Belzig2005a, Koch2005b, Aghassi2006a}.

After introducing our model, we review in
Sec.~\ref{sec:scattering} the Floquet scattering theory for the
computation of the current and the zero-frequency noise.  In
Sec.~\ref{sec:master}, we derive a master
equation approach which is applicable also in the presence of
electron-electron interactions.  These formalisms are in
Sec.~\ref{sec:pumping} employed for investigating the influence of
Coulomb repulsion on the noise in non-adiabatic electron pumps.

\subsection{Wire-lead model}
\label{sec:model}

\begin{SCfigure}[4][tb]
\includegraphics[width=0.45\textwidth]{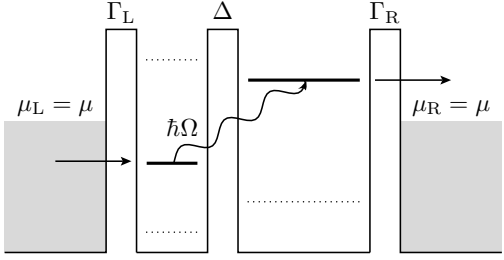}
\caption{\label{fig:levels} Level structure of a double quantum dot with
$N=2$ orbitals. The terminating sites are coupled to leads with chemical
potential $\mu_L$ and $\mu_\mathrm{R}=\mu_\mathrm{L}+eV$, respectively.}
\end{SCfigure}

A frequently used model for a nanoscale conductors like molecular
wires or coupled quantum dots is sketched in Fig.~\ref{fig:levels}.  It
is described by the time-dependent Hamiltonian
\begin{equation}
\label{eq:H}
H(t) = H_\mathrm{wire}(t) + H_\mathrm{leads} +
H_\mathrm{contacts},
\end{equation}
where the different terms correspond to the central conductor
(``wire''), electron reservoirs (``leads''), and the wire-lead
couplings, respectively.  We focus on the regime of coherent quantum
transport where the main physics at work occurs on the wire itself.
In doing so, we neglect other possible influences originating from
driving-induced hot electrons in the leads and dissipation on the
wire.  Then, the wire Hamiltonian reads in a tight-binding
approximation with $N$ orbitals $|n\rangle$
\begin{equation}
H_{\rm wire}(t)
= \sum_{n,n',s,s'} H_{nn'}(t) c^{\dag}_{ns} c^{\phantom{\dag}}_{n's'}
+ H_\text{interaction} .
\label{Hw}
\end{equation}
For a molecular wire, this constitutes the so-called H\"uckel
description where each site corresponds to one atom.  The fermion
operators $c_{ns}$, $c_{ns}^{\dag}$ annihilate and create,
respectively, an electron with spin $s={\uparrow},{\downarrow}$ in the
orbital $|n\rangle$.  The influence of an applied ac field or an
oscillating gate voltage with frequency $\Omega=2\pi/{\cal T}$ results
in a periodic time-dependence of the wire Hamiltonian:
$H_{nn'}(t+{\cal T})=H_{nn'}(t)$.
For the interaction Hamiltonian, we assume a capacitor model, so that
\begin{equation}
H_\text{interaction} = \frac{U}{2} N_\text{wire}(N_\text{wire}-1)
\end{equation}
where $N_\text{wire} = \sum_{ns} c_{ns}^\dagger c_{ns}$ describes the
number of
electrons on the wire. Below we shall focus on two limits, namely the
interaction-free case $U=0$ and strong interaction, $U\to\infty$,
which finally means that the Coulomb repulsion is so strong that only
states with zero or one excess electron play a role.

The leads are modelled by ideal electron gases,
\begin{equation}
H_\mathrm{leads}=\sum_{q,s} \epsilon_{q} (c^{\dag}_{\mathrm{L}qs}
c^{\phantom{\dag}}_{\mathrm{L}qs} + c^{\dag}_{\mathrm{R}qs} 
c^{\phantom{\dag}}_{\mathrm{R}qs}),
\label{eq:Hl}
\end{equation}
where $c_{\mathrm{L}q}^{\dag}$ ($c_{\mathrm{R}q}^{\dag}$) creates an electron
in the state $|\mathrm{L}q \rangle$ ($|\mathrm{R}q \rangle$) in the left
(right) lead.  The wire-lead tunnelling Hamiltonian
\begin{equation}
H_{\rm contacts} = \sum_{q,s} \left( V_{\mathrm{L}qs} c^{\dag}_{\mathrm{L}qs}
c^{\phantom{\dag}}_{1s}
+ V_{\mathrm{R}qs} c^{\dag}_{\mathrm{R}qs} c^{\phantom{\dag}}_{Ns}
\right) + \mathrm{h.c.}
\label{eq:Hc}
\end{equation}
establishes the contact between the sites $|1\rangle$, $|N\rangle$
and the respective lead.
This tunnelling coupling is described by the spectral density
\begin{equation}
\label{eq:Gamma}
\Gamma_\ell (\epsilon) = 2\pi \sum_q |V_{\ell q}|^2
\delta(\epsilon-{\epsilon}_q)
\end{equation}
of lead $\ell=\mathrm{L},\mathrm{R}$.
In the following, we restrict ourselves to the so-called wide-band
limit in which the spectral density is assumed to be
energy-independent, $\Gamma_\ell(\epsilon)\to\Gamma_\ell$.

To fully specify the dynamics, we choose as an initial condition
for the left/right lead a grand-canonical electron ensemble at
temperature $T$ and electro-chemical potential 
$\mu_{\mathrm{L}/\mathrm{R}}$.  Thus, the initial density matrix reads
\begin{equation}
\rho_0 \propto \e^{-(H_\mathrm{leads} -\mu_\mathrm{L} N_\mathrm{L}
-\mu_\mathrm{R} N_\mathrm{R})/k_\mathrm{B}T} ,
\label{ic}
\end{equation}
where $N_{\ell}=\sum_{qs} c^{\dag}_{\ell qs} c^{\phantom{\dag}}_{\ell qs}$
is the number of electrons in lead $\ell$ and $k_\mathrm{B} T$ denotes
the Boltzmann constant times temperature.  An applied voltage $V$ maps
to a chemical potential difference $\mu_\mathrm{R}-\mu_\mathrm{L}=eV$
with $-e$ being the electron charge.  Then, at initial time $t_0$, the
only nontrivial expectation values of the wire operators read
$
\langle c_{\ell'q's'}^\dagger c_{\ell q s}\rangle
= f_\ell(\epsilon_q) \delta_{\ell\ell'}\delta_{qq'} \delta_{ss'}
$
where $f_\ell(\epsilon)=(1+\exp[(\epsilon-\mu_\ell)/k_\mathrm{B}T])^{-1}$
denotes the Fermi function.

\subsection{Charge, current, and current fluctuations}

To avoid the explicit appearance of commutators in the definition of
correlation functions, we perform the derivation of the central transport
quantities in the Heisenberg picture.  As a starting point we choose the
operator
\begin{equation}
\label{eq:Q}
Q_\ell(t) = eN_\ell(t)-eN_\ell(t_0)
\end{equation}
which describes the charge accumulated in lead $\ell$ with respect to the
initial state.  Due to total charge conservation, $Q_\ell$ equals the net
charge transmitted across the contact $\ell$; its time derivative defines
the corresponding current
\begin{equation}
\label{eq:Ioperator}
I_\ell(t) = \frac{\d}{\d t} Q_\ell(t) .
\end{equation}
The current noise is described by the symmetrised correlation function
\begin{equation}
S_\ell (t,t')
= \frac{1}{2} \big\langle [\Delta I_\ell(t),\Delta I_\ell(t')]_+
\big\rangle
\label{eq:S}
\end{equation}
of the current fluctuation operator $\Delta I_\ell(t) = I_\ell(t)-\langle
I_\ell(t)\rangle$, where the anticommutator $[A,B]_+=AB+BA$ ensures
hermiticity.  At long times, $S_\ell(t,t') =
S_\ell(t+\mathcal{T},t'+\mathcal{T})$ shares the time-periodicity of
the driving \cite{Camalet2004a}.  Therefore, it is possible to
characterise the noise level by the zero-frequency component of
$S_\ell(t,t-\tau)$ averaged over the driving period,
\begin{equation}
\label{eq:barS}
\bar S_\ell = \frac{1}{\cal T} \int_0^{\cal T} \d t
\int_{-\infty}^\infty \d \tau\, S_\ell (t,t-\tau) .
\end{equation}
Moreover for two-terminal devices, $\bar S_\ell$ is
independent of the contact $\ell$, i.e., $\bar S_\mathrm{L}=\bar S_\mathrm{R}
\equiv \bar S$.

The evaluation of the zero-frequency noise $\bar S$ directly from its
definition \eqref{eq:barS} can be tedious due to the explicit appearance of
both times, $t$ and $t-\tau$.  This inconvenience can be circumvented by
employing the relation
\begin{equation}
\label{eq:Qdiff}
\frac{\d}{\d t}\Big( \langle Q_\ell^2(t)\rangle -\langle Q_\ell(t)\rangle^2\Big)
= 2\int_0^\infty \d\tau\, S_\ell(t,t-\tau) ,
\end{equation}
which follows from the integral representation of
Eqs.\ \eqref{eq:Q} and \eqref{eq:Ioperator},
$ Q_\ell(t) = \int_{t_0}^t \d t'\, I_\ell(t')$, in the limit $t_0\to-\infty$.
By averaging Eq.~\eqref{eq:Qdiff} over the driving period and using
$S(t,t-\tau) = S(t-\tau,t)$, we obtain
\begin{equation}
\label{eq:barSQ}
\bar S = \Big\langle \frac{\d}{\d t}\langle\Delta
Q_\ell^2(t)\rangle\Big\rangle_t \,,
\end{equation}
where $\Delta Q_\ell = Q_\ell-\langle Q_\ell\rangle$ denotes the charge
fluctuation operator and $\langle\ldots\rangle_t$ the time average.
The fact that the time average can be evaluated from the limit $\bar S =
\lim_{t_0\to -\infty} \langle \Delta Q_\ell^2(t)\rangle/(t-t_0) > 0$ allows
to interpret the zero-frequency noise as the ``charge diffusion
coefficient''.
As a dimensionless measure for the \textit{relative} noise strength, we
employ the so-called Fano factor \cite{Fano1947a}
\begin{equation}
\label{eq:Fano}
F = \frac{{\bar S}}{e|\bar I|} \, ,
\end{equation}
which can provide information about the nature of the transport
mechanism \cite{Blanter2000a, Grabert2002a}.
Here, $\bar I$ denotes the time-average of the current expectation value
$\langle I_\ell(t)\rangle$.
Historically, the zero-frequency noise \eqref{eq:barS} contains a
factor $2$, i.e.\ $\bar S'=2\bar S$, resulting from a different
definition of the Fourier transform.  Then, the Fano factor is defined
as $F=\bar S'/2e|\bar I|$.

\subsection{Full counting statistics}

A more complete picture of the current fluctuations beyond second
order correlations is provided by the full counting statistics. It
is determined by the moment generating function
\begin{equation}
\phi(\chi,t)
= \langle \e^{\i\chi N_\mathrm{L}} \rangle_t
\end{equation}
and allows the direct computation of the $k$th moment of the charge in
the left lead via the relation
\begin{equation}
\label{FCS_moment}
\langle Q_\mathrm{L}^k(t) \rangle
= e^k \frac{\partial^k\phi(\chi)}{\partial (\i\chi)^k}
  \phi(\chi,t) \Big|_{\chi=0} .
\end{equation}
Subtracting from the moments the trivial contributions that depend on
a shift of the initial values, one obtains the cumulants.  They are
defined and generated via the so-called cumulant generating function
$\ln\phi(\chi,t)$ which replaces $\phi$ in
Eq.~\eqref{FCS_moment} \cite{Risken}, so that the $k$th cumulant reads
\begin{equation}
C_k
= e^k \frac{\partial^k\phi(\chi)}{\partial (\i\chi)^k}
  \ln\phi(\chi,t) \Big|_{\chi=0} .
\end{equation}

In a continuum limit for the leads, both the moments and the cumulants
diverge as a function of time, and one focusses on the rates at which
these quantities change in the long-time limit. This establishes
between the first two cumulants and $I(t)$ and $S(t)$ the relations
\begin{align}
\label{FCS_I}
I(t) ={}&
-\i e\frac{\partial}{\partial\chi} \dot C(\chi,t)\Big|_{\chi=0},
\\
\label{FCS_S}
S(t) ={}&
-e^2 \frac{\partial^2}{\partial\chi^2} \dot C(\chi,t)\Big|_{\chi=0}.
\end{align}
For driven systems, these quantities are time-dependent even in the
asymptotic limit and, thus, we characterise the transport by the
corresponding averages over one driving period.  Then expressions
\eqref{FCS_I} and \eqref{FCS_S} become identical to the previously
defined time averages $\bar I$ and $\bar S$, respectively.
Herein we restrict ourselves to the computation of the first and
the second cumulant, despite the fact that also higher-order cumulants
can be measured \cite{Reulet2003a, Ankerhold2005a}.

\section{Floquet scattering theory}
\label{sec:scattering}

We now derive from the model described in Section~\ref{sec:model} in
the absence of electron-electron interactions expressions for both the
current through the wire and the associated noise by solving
the corresponding Heisenberg equations of motions.  Since for $U=0$,
the both spin directions contribute independently to the current, we
ignore the spin index which means that we consider the current per
spin projection.  We start from the equations of motion for the
annihilation operators in lead $\ell$,
\begin{equation}
\label{Heisenberg:lead}
\dot c_{\ell q}
= -\frac{\i}{\hbar} \epsilon_{\ell q} c_{\ell q} -
\frac{\i}{\hbar}V_{\ell q}\,c_{n_\ell} ,
\end{equation}
which are straightforwardly integrated to read
\begin{equation}
\label{c:lead}
c_{\ell q}(t)
=  c_{\ell q}(t_0)e^{-\i\epsilon_{\ell q}(t-t_0)/\hbar}
   -\frac{\i}{\hbar} V_{\ell q} \int_0^{t-t_0}\! \d\tau\,
   e^{-\i\epsilon_{\ell q}\tau/\hbar} c_{n_\ell}(t-\tau) ,
\end{equation}
where $n_\ell$ denotes the molecular wire site attached to lead $\ell$,
i.e., $n_\mathrm{L}=1$ and $n_\mathrm{R}=N$.  Inserting \eqref{c:lead} into the
Heisenberg equations for the wire operators yields in the asymptotic
limit $t_0\to-\infty$
\begin{align}
\label{c:xi:1N}
\dot c_{n_\ell}(t)
=& -\frac{\i}{\hbar} \sum_{n'} H_{n_\ell,n'}(t)\, c_{n'}(t)
   -\frac{1}{2\hbar}\Gamma_{\ell}\,
    c_{n_\ell}(t) +\xi_{\ell}(t) ,\\
\dot c_n(t)
=& -\frac{\i}{\hbar} \sum_{n'} H_{nn'}(t)\, c_{n'}(t)\,, \quad n=2,\ldots,N-1.
\label{c:xi:n}
\end{align}
For a energy-dependent spectral density $\Gamma_\ell =
\Gamma_\ell(\epsilon)$, the dissipative part of the Heisenberg equation
\eqref{c:xi:1N} acquires a memory kernel which complicates not only
its solution but also the derivation of a current formula. For
details, we refer the reader to Ref.~\cite{Kohler2005a}.

The influence of the operator-valued Gaussian noise
\begin{equation}
\label{xi}
\xi_{\ell}(t)
= -\frac{\i}{\hbar} \sum_q V^*_{\ell q}\,
  \e^{-\i\epsilon_{\ell q}(t-t_0)/\hbar} \, c_{\ell q}(t_0)
\end{equation}
is fully specified by the expectation values
$
\langle\xi_\ell(t)\rangle =  0$ and
\begin{equation}
\label{xi2}
\langle\xi^\dagger_{\ell'}(t')\, \xi_{\ell}(t)\rangle
 = \delta_{\ell\ell'}\frac{\Gamma_\ell}{2\pi\hbar^2} \int \d\epsilon\,
   \e^{-\i\epsilon(t-t')/\hbar}\,f_\ell(\epsilon) \, ,
\end{equation}
which follow directly
from the definition \eqref{xi} and the initial conditions~\eqref{ic}.
It is convenient to define the Fourier representation of the noise operator,
$\xi_\ell(\epsilon) = \int\d t\exp(\i\epsilon t/\hbar) \xi_\ell(t)$ whose
correlation function
\begin{equation}
\label{xi2.epsilon}
\langle\xi_\ell^\dagger(\epsilon) \,\xi_{\ell'}(\epsilon')\rangle
= 2\pi \Gamma_\ell \,f_\ell(\epsilon)\,
  \delta(\epsilon-\epsilon')\, \delta_{\ell\ell'}
\end{equation}
follows directly from Eq.~\eqref{xi2}.

\subsection{Retarded Green function}

The equations of motion \eqref{c:xi:1N} and \eqref{c:xi:n} represent a
set of linear inhomogeneous equations and, thus, can be solved with
the help of a retarded Green function $G(t,t')
=-(\i/\hbar)\,U(t,t')\theta(t-t')$, which obeys
\begin{equation}
\label{G.def}
\Big(\i\hbar\frac{\d}{\d t} - \mathcal{H}(t) + \frac{\i}{2}\Gamma\Big) G(t,t')
= \delta(t-t') ,
\end{equation}
where $\Gamma(t) = |1\rangle\Gamma_\mathrm{L}(t)\langle 1| +
|N\rangle\Gamma_\mathrm{R}(t)\langle N|$ and $\mathcal{H}(t)$ is the
one-particle Hamiltonian corresponding to Eq.~\eqref{Hw}.  At this stage, it
is important to note that the propagator of the homogeneous
equations obeys $U(t,t')=U(t+\T,t'+\T)$.  Accordingly, the Fourier
representation of the retarded Green function
\begin{equation}
\label{G(t,epsilon)}
G(t,\epsilon)
=  -\frac{\i}{\hbar} \int_0^\infty \! d\tau\, \e^{\i\epsilon\tau/\hbar}
    U(t,t-\tau)
= G(t+\mathcal{T},\epsilon)
= \sum_{k=-\infty}^\infty \e^{-\i k\Omega t} G^{(k)}(\epsilon)
\end{equation}
is also $\T$-periodic in the time argument, so that it can be
represented as a Fourier series.  Physically, the Fourier coefficients
$G^{(k)}(\epsilon)$ describe the propagation of an electron
with initial energy~$\epsilon$ under the absorption (emission) of $|k|$
photons for $k>0$ ($k<0$).
In the limiting case of a time-independent situation, all sideband
contributions with $k\neq 0$ vanish and $G(t,\epsilon)$ becomes
time-independent and identical to $G^{(0)}(\epsilon)$.
From the definition \eqref{G.def} of the Green function and its Fourier
representation \eqref{G(t,epsilon)}, it can be shown that the solution of
the Heisenberg equations \eqref{c:xi:1N}, \eqref{c:xi:n} reads
\begin{equation}
\label{cn.epsilon}
c_n(t) = \frac{\i}{2\pi} \sum_\ell\int\d\epsilon\, \e^{-\i\epsilon t/\hbar}
G_{n,n_\ell}(t,\epsilon)\, \xi_\ell(\epsilon),
\end{equation}
where we have defined $G_{n,n_\ell}(t,\epsilon) = \langle n|G(t, \epsilon)|n_\ell\rangle$.

Below, we need for the elimination of back-scattering terms the relation
\begin{equation}
\label{datta}
G^\dagger(t,\epsilon')-G(t,\epsilon)
= \Big( \i\hbar \frac{\d}{\d t} - \epsilon' + \epsilon +\i\Gamma \Big)
  G^\dagger(t,\epsilon') G(t,\epsilon) .
\end{equation}
A proof starts from the definition of the Green function,
Eq.~\eqref{G.def}.  By Fourier transformation with respect to $t'$, we
obtain the relation
\begin{equation}
\label{G:epsilon:def}
\Big(\i\hbar\frac{\d}{\d t} + \epsilon - \mathcal{H}(t)
     +\frac{\i}{2}\Gamma \Big) G(t,\epsilon)
= \mathbf{1} ,
\end{equation}
which we multiply by $G^\dagger(t,\epsilon)$ from the left.  The difference
between the resulting expression and its hermitian adjoint with $\epsilon$
and $\epsilon'$ interchanged is relation~\eqref{datta}.

\subsection{Current through the driven molecular wire}

Owing to charge conservation, the (net) current flowing from lead $\ell$
into the molecular wire is determined by the negative time derivative of
the charge in lead $\ell$.  Thus, the current operator reads $I_\ell = \i
e[H(t),N_\ell]/\hbar$, where $N_\ell = \sum_q c_{\ell q}^\dagger c_{\ell
q}$ denotes the corresponding electron number and $-e$ the electron charge.
From Eqs.\ \eqref{c:lead} and \eqref{xi} then follows
\begin{equation}
\label{I.operator}
I_\mathrm{L}(t)
= \frac{e \Gamma_\mathrm{L}}{\hbar} \int_0^\infty \d\tau\,\big\{
   c_1^\dagger(t)c_1(t-\tau)
   +c_1^\dagger(t-\tau) c_1(t) \big\}
 -e\big\{c_1^\dagger(t)\xi_\mathrm{L}(t)+\xi_\mathrm{L}^\dagger(t)c_1(t)\big\}.
\end{equation}
This operator-valued expression for the time-dependent current is a convenient
starting point for the evaluation of expectation values like the dc
and ac current and the current noise.

\subsubsection{Time-average current}

To obtain the current $\langle I_\mathrm{L}(t)\rangle$, we insert the solution
\eqref{cn.epsilon} of the Heisenberg equation into the current operator
\eqref{I.operator} and use the expectation values \eqref{xi2.epsilon}.  The
resulting expression
\begin{equation}
\label{I.operator.xi}
\langle I_\mathrm{L}(t)\rangle
=  \frac{e\Gamma_\mathrm{L}}{h} \sum_\ell \Gamma_\ell \Im \int\d\epsilon
       f_\ell(\epsilon) \e^{\i\epsilon\tau/\hbar}
   G_{1\ell}^*(t,\epsilon)\, G_{1\ell}(t-\tau,\epsilon)
 + 2 e \Gamma_\mathrm{L}\Im \int\d\epsilon\,
     f_\mathrm{L}(\epsilon)
     G_{11}(t,\epsilon)
\end{equation}
still contains back-scattering terms $G_{11}$ and, thus, is not of a
``scattering form''.  Indeed, bringing \eqref{I.operator.xi} into a form
that resembles the static current formula requires
some tedious algebra.  Such a derivation has been presented for the linear
conductance of time-independent systems \cite{Fisher1981a}, for tunnelling
barriers \cite{Caroli1971a} and mesoscopic conductors \cite{Meir1992a} in
the static case for finite voltage, and for a wire consisting of levels
that couple equally strong to both leads \cite{Jauho1994a}.  For the
periodically time-dependent case in the absence of electron-electron
interactions, such an expression has been \textit{derived} only recently
\cite{Camalet2003a, Camalet2004a}.

Inserting the matrix element $\langle 1|\ldots|1\rangle$ of equation
\eqref{datta} eliminates the back-scattering terms and yields for the
time-dependent current the expression
\begin{equation}
\label{I(t)scattering}
\langle I_\mathrm{L}(t)\rangle
=  \frac{e}{h}  \int \d\epsilon\,
   \big\{ T_{\mathrm{L}\mathrm{R}} (t,\epsilon) f_\mathrm{R} (\epsilon)
        - T_{\mathrm{R}\mathrm{L}} (t,\epsilon) f_\mathrm{L} (\epsilon) \big\}
   - \frac{\d}{\d t} q_\mathrm{L}(t)
\end{equation}
where $q_\mathrm{L}(t)$
denotes the charge oscillating between the left lead and the wire.
Obviously, since $q_\mathrm{L}(t)$ is time-periodic and bounded, its
time derivative cannot contribute to the average current.  The
time-dependent current is determined by the time-dependent transmission
\begin{equation}
\label{T.LR(t)}
T_{\mathrm{L}\mathrm{R}}(t,\epsilon)
= \Gamma_\mathrm{L}\Gamma_\mathrm{R}
  \Re \int_0^\infty \!\d\tau\, \e^{\i\epsilon\tau/\hbar}
  G_{1N}^*(t,\epsilon)\, G_{1N}(t-\tau,\epsilon) .
\end{equation}
The corresponding expression for
$T_{\mathrm{R}\mathrm{L}}(t,\epsilon)$ follows from the replacement
$(L,1)\leftrightarrow(R,N)$.  We emphasise that expression
\eqref{I(t)scattering} obeys the form of the current
formula for a static conductor within
a scattering formalism.  In particular, consistent with
Refs.~\cite{Datta1992a,Datta1995a}, no ``Pauli blocking factors''
$(1-f_\ell)$ appear in our derivation.

The dc current obtained from \eqref{I(t)scattering} by time-averaging can be
written in an even more compact form if we insert for the Green
function the Fourier representation \eqref{G(t,epsilon)}. This results
in
\begin{equation}
\label{barI}
\bar I
= \frac{e}{h}  \sum_k \int\d\epsilon \left\{ T_{\mathrm{L}\mathrm{R}}^{(k)} (\epsilon)
  f_\mathrm{R} (\epsilon) - T_{\mathrm{R}\mathrm{L}}^{(k)} (\epsilon) f_\mathrm{L} (\epsilon) \right\} ,
\end{equation}
where
\begin{align}
\label{TLR}
T_{\mathrm{L}\mathrm{R}}^{(k)}(\epsilon)
=& \Gamma_\mathrm{L} \Gamma_\mathrm{R}
   \big|G_{1N}^{(k)}(\epsilon) \big|^2 , \\
\label{TRL}
T_{\mathrm{R}\mathrm{L}}^{(k)}(\epsilon)
=& \Gamma_\mathrm{R} \Gamma_\mathrm{L}
   \big|G_{N1}^{(k)}(\epsilon) \big|^2 ,
\end{align}
denote the transmission probabilities for electrons from the right to
the left lead and vice versa, respectively, with initial energy
$\epsilon$ and final energy $\epsilon+k\hbar\Omega$, i.e., the
probability for an scattering event under the absorption (emission) of
$|k|$ photons if $k>0$ ($k<0$).

For a static situation, all contributions with
$k\neq 0$ vanish and $T_{\mathrm{L}\mathrm{R}}^{(0)}(\epsilon) =
T_{\mathrm{R}\mathrm{L}}^{(0)}(\epsilon)$.  Therefore, it is possible
to write the current \eqref{barI} as a product of a single
transmission $T(\epsilon)$, which is independent of the direction, and
the difference of the Fermi functions,
$f_\mathrm{R}(\epsilon)-f_\mathrm{L}(\epsilon)$.  We emphasise that in
the driven case this no longer holds true.

\subsubsection{Noise power}

In order to derive a related expression for the time-averaged
current-current correlation function \eqref{eq:barS}, we insert the
current operator \eqref{I.operator} and the solution
\eqref{cn.epsilon} of the Heisenberg equations of motion.  Then, we
again employ relation \eqref{datta} and the shorthand notation
$\epsilon_k=\epsilon+k\hbar\Omega$, so that we finally obtain
\begin{equation}
\label{barS}
\begin{split}
\bar S
={}& \frac{e^2}{h} \sum_k \int \d\epsilon \Big\{
    \Gamma_\mathrm{R} \Gamma_\mathrm{R}
    \Big| \sum_{k'} \Gamma_\mathrm{L} ( \epsilon_{k'})
    G^{(k'-k)}_{1N}(\epsilon_k) \big[G_{1N}^{(k')}(\epsilon)\big]^* \Big|^2
    f_\mathrm{R} (\epsilon) \bar f_\mathrm{R} (\epsilon_k)
\\
&+  \Gamma_\mathrm{R} \Gamma_\mathrm{L}
    \Big| \sum_{k'}
    \Gamma_\mathrm{L} G^{(k'-k)}_{1N}(\epsilon_k)
    \big[ G_{11}^{(k')}(\epsilon) \big]^* - \i G^{(-k)}_{1N}(\epsilon_k)
    \Big|^2 f_\mathrm{L}(\epsilon) \bar f_\mathrm{R}(\epsilon_k)
    \Big\} \\
&+  \text{same terms with the replacement $(L,1) \leftrightarrow (R,N)$} .
\end{split}
\end{equation}

\subsubsection{Floquet decomposition in the wide-band limit}

Solving the equations of motion \eqref{G.def} for the Green function
is equivalent to computing a complete set of solutions for the
equation
\begin{equation}
\label{preFloquet}
\i\hbar\frac{\d}{\d t} |\psi(t)\rangle
= \Big(\mathcal{H}_\mathrm{wire}(t) - \frac{\i}{2}\Gamma \Big)|\psi(t)\rangle ,
\end{equation}
which is linear and possesses time-dependent,
$\mathcal{T}$-periodic coefficients. Thus, it is possible to construct
a complete set of solutions with the Floquet ansatz
\begin{align}
|\psi_\alpha(t)\rangle
= & \exp[(-\i\epsilon_\alpha/\hbar-\gamma_\alpha)t] |u_\alpha(t)\rangle
,
\\
|u_{\alpha}(t)\rangle
= & \sum_k \exp(-\i k\Omega t)|u_{\alpha,k}\rangle .
\end{align}
The so-called Floquet states $|u_{\alpha}(t)\rangle$ obey the
time-periodicity of $\mathcal{H}_\mathrm{wire}(t)$ and have been
decomposed into a Fourier series.  In a Hilbert space that is extended
by a periodic time coordinate, the so-called Sambe space
\cite{Sambe1973a}, they obey the Floquet eigenvalue equation
\cite{Grifoni1998a, Buchleitner2002a}
\begin{equation}
\Big(\mathcal{H}_\mathrm{wire}(t) - \i\Sigma
-\i\hbar\frac{\d}{\d t}\Big)|u_{\alpha}(t)\rangle
= (\epsilon_{\alpha} -  \i\hbar\gamma_{\alpha}) |u_{\alpha}(t)\rangle .
\label{Fs}
\end{equation}
Due to the Brillouin zone structure of the Floquet spectrum
\cite{Shirley1965a,Sambe1973a,Grifoni1998a}, it is sufficient to
compute
all eigenvalues of the first Brillouin zone,
$-\hbar\Omega/2<\epsilon_\alpha \le \hbar\Omega/2$.  Since the
operator on
the l.h.s.\ of Eq.~\eqref{Fs} is non-Hermitian, the eigenvalues
$\epsilon_{\alpha} - \i\hbar\gamma_{\alpha}$ are generally complex-valued
and the (right) eigenvectors are not mutually orthogonal.  Thus, to
determine the propagator, we need to solve also the adjoint Floquet
equation yielding again the same eigenvalues but providing the adjoint
eigenvectors $|u_\alpha^+(t)\rangle$.
It can be shown that the Floquet states $|u_\alpha(t)\rangle$ together
with
the adjoint states $|u_\alpha^+(t)\rangle$ form at equal times a
complete
bi-orthogonal basis: $\langle u^+_{\alpha}(t)|u_{\beta}(t)\rangle =
\delta_{\alpha\beta}$ and $\sum_{\alpha} |u_{\alpha}(t)\rangle \langle
u^+_{\alpha} (t)|= \mathbf{1}$.  A proof requires to account for the
time-periodicity of the Floquet states since the eigenvalue equation
\eqref{Fs} holds in a Hilbert space extended by a periodic time
coordinate \cite{Jung1990a,Grifoni1998a}.

Using the Floquet equation \eqref{Fs}, it is straightforward to show
that the propagator can be written as
\begin{equation}
\label{Gtt}
U(t,t')
= \sum_\alpha \e^{-\i(\epsilon_\alpha/\hbar-\i\gamma_\alpha)(t-t')}
  |u_\alpha(t)\rangle\langle u^+_\alpha(t')| ,
\end{equation}
where the sum runs over all Floquet states within one Brillouin zone.
Consequently, the Fourier coefficients of the Green function read
\begin{align}
G^{(k)}(\epsilon)
=& -\frac{\i}{\hbar}\int_0^\T \frac{\d t}{\T} e^{\i k\Omega t}
   \int_0^\infty\!\! \d\tau
   \e^{\i\epsilon\tau/\hbar} U(t,t-\tau)
\\
=& \sum_{\alpha,k'}
   \frac{|u_{\alpha,k'+k}\rangle\langle u_{\alpha,k'}^+|}
      {\epsilon-(\epsilon_\alpha+k'\hbar\Omega-\i\hbar\gamma_\alpha)} .
\label{G}
\end{align}

In general, the Floquet equation \eqref{Fs} has to be solved
numerically.
In the zero
temperature limit considered here, the Fermi functions in the
expressions for the average current \eqref{barI} and the
zero-frequency noise \eqref{barS} become step functions.  Therefore,
the integrands are rational functions and the remaining energy
integrals can be performed analytically.

\section{Master equation approach}
\label{sec:master}

In the presence of electron-electron interactions, an exact treatment
of the electron transport within a scattering theory is no longer
possible and a master equation formalism can be an appropriate tool
for the computation of currents \cite{Gurvitz1996a, Stoof1996a,
Brune1997a, Lehmann2004a, Kohler2005a}.  Recently, master equations
have been established for the computation of current noise of various
static conductors as well \cite{Elattari2002a, Bagrets2003a,
Kiesslich2003a, Novotny2004a, Flindt2004a, Belzig2005a, Koch2005b,
Aghassi2006a}.  In the following, we develop such an approach for the
case of periodically time-dependent conductors.

\subsection{Perturbation theory and reduced density operator}

We start our derivation of a master equation formalism from the
Liouville-von Neumann equation $\i \hbar
\dot{\mathcal{R}}(t)=[H(t),\mathcal{R}(t)]$ for the total density
operator $\mathcal{R}(t)$.  By standard techniques we obtain the exact
equation of motion
\begin{equation}
\label{MEbasic}
 \frac{\d}{\d t} \widetilde{\mathcal{R}}(t) = 
 - \frac{\i}{\hbar} [\widetilde{H}_\mathrm{wire-leads}(t), \mathcal{R}(0)]
 - \frac{1}{\hbar^2}\int_0^{\infty}\d\tau [H_\mathrm{wire-leads},
   [\widetilde H_\mathrm{wire-leads}(t-\tau,t),\mathcal{R}(t)]] ,
\end{equation}
where the tilde denotes the interaction picture with respect to the
lead and the wire Hamiltonian,
$\tilde X(t,t') = U_0^\dagger(t,t') X U_0(t,t')$, and $U_0$ is the
propagator without the coupling. Below we will employ Floquet theory
in order to obtain explicit expressions for these operators.

As already discussed above, the moment generating function $\phi(\chi)
= \langle \exp(\i\chi N_\mathrm{L})\rangle$ contains the full
information about the counting statistics.  For its explicit
computation, we define in the Hilbert space of the wire the operator
\begin{equation}
\mathcal{F}(\chi,t) =\tr_\mathrm{leads} \{\e^{\i\chi N_\L} 
\mathcal{R}(t)\} ,
\end{equation}
whose limit $\chi\to 0$ obviously is the reduced density operator of
the wire, $\mathcal{F}(0,t) = \rho$.  After tracing out the wire
degrees of freedom, $\mathcal{F}$ becomes the moment generating
function $\phi(\chi) = \tr_\mathrm{wire}\mathcal{F}$.  It will prove
convenient to decompose $\mathcal{F}$ into a Taylor series,
\begin{equation}
\mathcal{F}
= \rho + \sum_{k=1}^\infty \frac{(\i\chi)^k}{k!} \mathcal{F}_k ,
\end{equation}
where the coefficients $\mathcal{F}_k = \tr_\text{leads}
(N_\mathrm{L}^k \mathcal{R})$ provide direct access to the moments
$\langle N_\mathrm{L}^k\rangle = \tr_\text{wire} \mathcal{F}_k$.

Our strategy is now to derive from the master equation \eqref{MEbasic}
for the full density operator an equation of motion for the
$\mathcal{F}_k$.  For that purpose, we transform the master equation
for $\tilde{\mathcal{R}}$ back to the Schr\"odinger picture and multiply
it from the left by the operator $\exp(\i\chi N_\mathrm{L})$.
By tracing out the lead degrees of freedom and using the commutation
relations $[N_\L , V]=V$ and $[N_\L , V^\dagger]=-V^\dagger$, we obtain
\begin{equation}
\label{eq:motion_F}
\frac{\d}{\d t}{\mathcal{F}}(\chi,t)
= \{\mathcal{L} + (\e^{ \i \chi} -1)\mathcal{J}_+ +
(\e^{- \i \chi} -1)\mathcal{J}_-\}\mathcal{F}(\chi,t) .
\end{equation}
In order to achieve this compact notation, we have defined the
superoperators $\mathcal{J}_\pm$ and the time-dependent
Liouville operator
\begin{equation}
\label{L}
\begin{aligned}
\mathcal{L}(t)\, X  ={}&
-\frac{\i}{\hbar} [H_\text{wire}(t), X]
\\
&+
\frac{\Gamma_\L}{2 \pi} \int_0^\infty \d \tau \int \d \epsilon
 \Big[ \e^{\i \epsilon \tau} \big(
-c_1 \tilde{c}_1^\dagger  X  f_\L(\epsilon)
+\tilde{c}_1^\dagger  X  c_1 f_\L(\epsilon)
- X  \tilde{c}_1^\dagger c_1 \bar{f}_\L(\epsilon)
+ c_1  X  \tilde{c}_1^\dagger \bar{f}_\L(\epsilon)
\big)
\\
& \hspace{10ex}+\e^{-\i \epsilon \tau} \big(
- X  \tilde{c}_1 c_1^\dagger f_\L(\epsilon)
+c_1^\dagger  X  \tilde{c}_1 f_\L(\epsilon)
-c_1^\dagger \tilde{c}_1  X  \bar{f}_\L(\epsilon)
+ \tilde{c}_1  X  c_1^\dagger \bar{f}_\L(\epsilon)
\big)
\Big]
\\
&{}+ \text{same terms with the replacement $1,\L \rightarrow N, \R$} , 
\end{aligned}
\end{equation}
which also determines the time-evolution of the reduced density operator,
$\dot{\rho} = \mathcal{L}(t)\,\rho$.
The tilde denotes the interaction picture operator $\tilde c =
\tilde c(t,t-\tau)$ and $f_\ell$ the Fermi function of lead $\ell$,
while $\bar f_\ell = 1-f_\ell$.
The current operators
\begin{align}
\mathcal{J_+}(t)\, X  ={}& 
\frac{\Gamma_\L}{2 \pi} \int_0^\infty \d \tau \int \d \epsilon
\big( \e^{\i \epsilon \tau} \tilde{c}_1^\dagger X c_1
+ \e^{-\i \epsilon \tau} c_1^\dagger  X  \tilde{c}_1
\big) f_\L(\epsilon) ,
\\
\mathcal{J_-}(t)\, X  ={}& 
\frac{\Gamma_\L}{2 \pi} \int_0^\infty \d \tau \int \d \epsilon
\big(  \e^{\i \epsilon \tau}
c_1 X \tilde{c}_1^\dagger   +
\e^{-\i \epsilon \tau} \tilde{c}_1  X  c_1^\dagger
\big) \bar{f}_\L(\epsilon) ,
\end{align}
describe the tunnelling of an electron from the left lead to the wire
and the opposite process, respectively.  Note that these
superoperators still contain a non-trivial time-dependence stemming
from the interaction-picture representation of the creation and
annihilation operators of wire electrons.

\subsection{Computation of moments and cumulants}

For computation of the current \eqref{FCS_I} and the zero-frequency
noise \eqref{FCS_S}, we generalise the approach of
Ref.~\cite{Novotny2004a} to the time-dependent case.  Since we
restrict the noise characterisation to the Fano factor, it is
sufficient to compute the long-time behaviour of the first and the
second moment of the electron number in the left lead.  This
information is fully contained in the time-derivative of the operator
$\mathcal{F}$ up to second order in $\chi$, for which we obtain by Taylor
expansion of the equation of motion \eqref{eq:motion_F} the hierarchy
\begin{align}
\label{ME-rho}
\dot\rho
={}& \mathcal{L}(t)\,\rho ,
\\
\label{ME-sigma1}
\dot{\mathcal{F}}_1
={}& \mathcal{L}(t)\,\mathcal{F}_1 + 
\big( \mathcal{J}_+(t) + \mathcal{J}_-(t) \big) \rho ,
\\
\label{ME-sigma2}
\dot{\mathcal{F}}_2
={}& \mathcal{L}(t)\,\mathcal{F}_2 + 
2\big( \mathcal{J}_+(t) + \mathcal{J}_-(t) \big) \mathcal{F}_1 +
\big( \mathcal{J}_+(t) - \mathcal{J}_-(t) \big) \rho .
\end{align}
The first equation determines the time-evolution of the reduced
density operator, which in the long-time limit becomes the stationary
solution $\rho_0(t)$.  Note that for a driven system, it generally is
time-dependent.  Replacing in Eq.~\eqref{ME-sigma1} $\rho$ by $\rho_0$
and using the fact that $\tr_\text{wire}\mathcal{L}X = 0$ for any
operator $X$, we obtain the stationary current
\begin{equation}
\label{I(t)ME}
I(t)
= -e\tr_\text{wire}\dot{\mathcal{F}}_1
= -e\tr_\text{wire}(\mathcal{J}_+ +\mathcal{J}_-)\rho_0(t) .
\end{equation}
The dc current follows simply by averaging over one driving period and
one ends up with the current formula of Ref.~\cite{Kaiser2006b}.

The computation of $\mathcal{F}_1(t)$ is hindered by the fact that
the inverse of a Liouvillian generally does not exist.  For static
systems this is obvious from the fact that the stationary solution fulfils
$\mathcal{L}\rho_0 = 0$, which implies that $\mathcal{L}$ is singular.
This unfortunately also complicates the computation of the second
cumulant and we proceed in the following way:
We start from Eq.~\eqref{eq:Qdiff} which relates the zero frequency
noise to the charge fluctuation in the leads and write the time
derivative of the first and the second moment of the electron number
in the left lead by the operators $\dot{\mathcal{F}}_{1,2}$. From the
equations of motion \eqref{ME-sigma1} and \eqref{ME-sigma2}, we then
find
$S = e^2\tr_\text{wire}\{ 2(\mathcal{J}_++\mathcal{J}_- -I)\mathcal{F}_1
- (\mathcal{J}_+ -\mathcal{J}_-)\rho \}$, where we again used the
relation $\tr_\text{wire}\mathcal{L}X = 0$.
An important observation is now that the first part of this expression
vanishes for $\mathcal{F}_1 \propto \rho_0$, which can easily be
demonstrated by inserting the current expectation value~\eqref{I(t)ME}.
Since $\rho_0 \tr_\text{wire}$ acts as a projector onto the stationary
solution $\rho_0$, we can define the ``perpendicular'' part
\begin{equation}
\label{Fperp}
\mathcal{F}_{1\perp} = \mathcal{F}_{1} - \rho_0\tr_\text{wire}\mathcal{F}_1,
\end{equation}
which fulfils the relation $\tr_\text{wire}\mathcal{F}_{1\perp} = 0$
and obeys the equation of motion
\begin{equation}
\label{eomFperp}
\dot{\mathcal{F}}_{1\perp}
= \mathcal{L}(t)\,\mathcal{F}_{1\perp} + 
\big( \mathcal{J}_+(t) + \mathcal{J}_-(t) -I(t) \big) \rho_0(t) ,
\end{equation}
We will see below that in contrast to $\mathcal{F}_1$, the long-time
limit of the traceless $\mathcal{F}_{1\perp}$ can be computed directly
from the equation of motion \eqref{eomFperp}.
Upon inserting Eq.~\eqref{Fperp} into the equation of motion
\eqref{ME-sigma2}, we finally obtain for the still
time-dependent ``charge diffusion coefficient'' the expression
\begin{equation}
\label{Soperator}
S(t) = e^2\tr_\text{wire}\big\{
    2(\mathcal{J}_++\mathcal{J}_-)\mathcal{F}_{1\perp}
    -(\mathcal{J}_+-\mathcal{J}_-)\rho_0 \big\} ,
\end{equation}
whose time-average finally provides the Fano factor $F=\bar S/e\bar I$.

\subsection{Floquet decomposition}

The remaining task is now to compute the stationary solutions
$\rho_0(t)$ and $\mathcal{F}_{1\perp}(t)$ from the time-dependent
equations of motion \eqref{ME-rho} and \eqref{ME-sigma1}.  Like for
the computation of the dc current in our previous work
\cite{Kaiser2006b}, we solve this problem within a Floquet treatment
of the isolated wire, which provides a convenient representation of
the electron creation and annihilation operators.

\subsubsection{Fermionic Floquet operators}

In the driven wire Hamiltonian \eqref{Hw}, the single-particle
contribution commutes with the interaction term and, thus, these two
Hamiltonians possess a complete set of common many-particle
eigenstates.  Here we start by diagonalising the first part
of the Hamiltonian which describes the single-particle dynamics
determined by the time-periodic matrix elements $H_{nn'}(t)$.
According to the Floquet theorem, the corresponding (single particle)
Schr\"odinger equation possesses a complete solution of the form
\begin{equation}
    |\Psi_{\alpha}(t)\rangle
    =
    \e^{-\i \epsilon_{\alpha} t / \hbar} 
    |\varphi_{\alpha}(t)\rangle,
\end{equation}
with the so-called quasienergies $\epsilon_\alpha$ and the
$\mathcal{T}$-periodic Floquet states
\begin{equation}
    |\varphi_{\alpha}(t)\rangle =  \sum_{k} \e^{- \i k \Omega t}
    |\varphi_{\alpha,k}\rangle.
\end{equation}
The Floquet states and the quasienergies are obtained by solving the
eigenvalue problem
\begin{equation}
    \label{eq:quasienergy}
    \Big( \sum_{n,n'} \rangle n| H_{nn'}(t) \langle n'|  -\i \hbar
    \frac{\text{d}}{\text{d}t} \Big) |\varphi_{\alpha}(t)\rangle
  = \epsilon_{\alpha} |\varphi_{\alpha}(t)\rangle ,
\end{equation}
whose solution allows one to construct via Slater determinants
many-particle Floquet states.
In analogy to the quasimomenta in Bloch theory for spatially periodic
potentials, the quasienergies $\epsilon_{\alpha}$ come in classes
\begin{equation}
    \epsilon_{\alpha,k}=\epsilon_\alpha + k\hbar \Omega,
    \quad k \in  \mathbb{Z},
\end{equation}
of which all members represent the same physical solution of the Schr\"odinger
equation. Thus we can restrict ourselves to states within one Brillouin
zone like for example $0 \leq \epsilon_\alpha < \hbar \Omega$.

For the numerical computation of the operators $\rho_0$ and
$\mathcal{F}_{1\perp}$, it is essential to have an explicit expression for
the interaction picture representation of the wire operators.  It can
be obtained from the (fermionic) Floquet creation and annihilation
operators \cite{Kohler2005a} defined via the transformation
\begin{equation}
    \label{c.alpha}
    c_{\alpha s} (t) = \sum_n \langle\varphi_{\alpha}(t)| n\rangle c_{ns} .
\end{equation}
The inverse transformation
\begin{equation}
    \label{eq:floquet_operator}
    c_{ns} = \sum_\alpha \langle n |\varphi_{\alpha}(t)\rangle c_{\alpha s}(t)
\end{equation}
follows from the mutual orthogonality and the completeness of the
Floquet states at equal times \cite{Grifoni1998a}.  Note that the
right-hand side of Eq.~\eqref{eq:floquet_operator} becomes time
independent after the summation.
The Floquet annihilation operator \eqref{c.alpha} has the interaction
picture representation
\begin{eqnarray}
  \tilde{c}_{\alpha s}(t,t')
  &=& U^\dagger_0 (t,t')\, c_{\alpha s}(t)\,  U^{\vphantom{\dagger}}_0 (t,t')
    \\
  &=& \e^{- \i(\epsilon_{\alpha}+U N_\text{wire}) (t-t') / \hbar}
    c_{\alpha s}(t') ,
  \label{c.alpha.interaction}
\end{eqnarray}
with the important feature that the time difference $t-t'$ enters
only via the exponential prefactor.  This allows us to evaluate
the $\tau$-integration of the master equation \eqref{ME-rho}
after a Floquet decomposition.
Relation \eqref{c.alpha.interaction} can easily be shown by computing
the time derivative with respect to $t$ which by use of the Floquet
equation \eqref{eq:quasienergy} becomes
\begin{equation}
\label{c-t-t'}
   \frac{\d}{\d t} \tilde c_{\alpha s}(t,t')
 = -\frac{\i}{\hbar} (\epsilon_\alpha + UN_\text{wire})\,
   \tilde c_{\alpha s}(t,t') .
\end{equation}
Together with the initial condition $\tilde c_\alpha(t',t') =
c_\alpha(t')$ follows relation \eqref{c.alpha.interaction}.
Note that the time evolution induced by $\mathcal{H}_\text{wire}(t)$
conserves the number of electrons on the wire.

\subsubsection{Master equation and current formula}

In order to make use of the Floquet ansatz, we decompose the master
equation \eqref{ME-rho} and the
current formula \eqref{I(t)ME} into the Floquet basis derived
in the last subsection.  For that purpose we use the fact that
we are finally interested in the current at asymptotically large times
in the limit of large interaction $U$.
The latter has the consequence that only wire states with at most one
excess electron play a role, so that the stationary density operator
$\rho_0(t)$ can be decomposed into the $2N+1$ dimensional basis
$\{ |0\rangle, c_{\alpha s}^\dagger(t)\,|0\rangle \}$, where
$|0\rangle$ denotes the wire state in the absence of excess electrons
and $s=\uparrow,\downarrow$.
Moreover, it can be shown that at large times, the density operator
becomes diagonal in the electron number $N_\text{wire}$, so that
a proper ansatz reads
\begin{equation}
  \label{eq:decomposition}
  \rho_0(t)
  = |0\rangle \rho_{00}(t) \langle 0|
    + \sum_{\alpha,\beta,s,s'} c^\dagger_{\alpha s}
    |0\rangle \rho_{\alpha s,\beta s'}(t)\langle0| c_{\beta s'} .
\end{equation}
Note that we keep terms with $\alpha\neq\beta$, which means that we
work beyond a rotating-wave approximation.  Indeed in a
non-equilibrium situation, the off-diagonal density matrix elements
$\rho_{\alpha\beta}$ will not vanish and neglecting them might lead to
artefacts \cite{Novotny2002a, Kohler2005a}.

By inserting the decomposition \eqref{eq:decomposition}
into the master equation \eqref{ME-rho}, we obtain an equation
of motion for the matrix elements $\rho_{\alpha s,\beta s'}
= \langle0|c_{\alpha s}\rho_\text{wire} c^\dagger_{\beta s'}|0\rangle$.
We evaluate the trace over the lead states and compute the
matrix element $\langle 0|c_{\alpha s}(t)\ldots c_{\beta s'}^\dagger(t)|0\rangle$.
Thereby we neglect the two-particle terms which are of the structure
$c_{\alpha s}^\dagger c_{\beta s}^\dagger |0\rangle\langle 0| c_{\beta s}
c_{\alpha s}$.  Formally, these terms drop out in the limit of strong
Coulomb repulsion because they are accompanied by a rapidly oscillating
phase factor $\exp({-\i
UN_\text{wire}\tau}/\hbar)$.  Then the $\tau$-integration results in a
factor $f_\L(\epsilon_{\alpha,k}+U)$ which vanishes in the limit of
large $U$.  Since the total Hamiltonian~\eqref{eq:H} is diagonal in
the spin index $s$, we find that the density matrix elements
$\rho_{\alpha s, \beta s'}$ are spin-independent as well so that after
a transient stage
\begin{equation}
  \rho_{\alpha {\uparrow},\beta {\uparrow}}(t)
  =\rho_{\alpha {\downarrow},\beta {\downarrow}}(t)
  \equiv \rho_{\alpha\beta}(t)
\end{equation}
and $\rho_{\alpha {\uparrow},\beta {\downarrow}} =0$.
Moreover, the stationary density operator \eqref{eq:decomposition}
obeys the time periodicity of the driving field \cite{Kohler2005a}
and, thus, can be decomposed into the Fourier series
\begin{equation}
\label{rho-fourier}
\rho_{\alpha\beta}(t) = \sum_k \e^{-\i k\Omega t} \rho_{\alpha\beta,k}
\end{equation}
and $\rho_{00}(t)$ accordingly.

After some algebra, we arrive at a set of $N^2$ coupled equations of
motion for $\rho_{\alpha\beta}(t)$ which in Fourier representation read
\begin{equation}
\label{eq:masterfourier}
    \begin{split}
    \i( \epsilon_{\alpha}-\epsilon_{\beta}-k\hbar\Omega)
    \rho_{\alpha \beta,k}
   =&\frac{\Gamma_\L }{2}\sum_{k',k''}\,
    \langle \varphi_{\alpha,k'+k''} | 1 \rangle 
    \langle 1 | \varphi_{\beta, k+k''} \rangle
    \rho_{00,k'}
    \big( f_\L(\epsilon_{\alpha,k'+k''})
    +f_\L(\epsilon_{\beta,k+k''}) \big)
    \\
   &-\frac{\Gamma_\L }{2}\sum_{\alpha',k',k''}
    \langle \varphi_{\alpha,k'+k''} | 1 \rangle
    \langle 1 | \varphi_{\alpha',k+k''} \rangle
    \rho_{\alpha'\beta, k'}\,
    \bar{f}(\epsilon_{\alpha',k+k''})
    \\
   &-\frac{\Gamma_\L }{2}\sum_{\beta',k',k''}
    \langle \varphi_{\beta',k'+k''} | 1 \rangle
    \langle 1 | \varphi_{\beta,k+k''} \rangle
    \rho_{\alpha\beta', k'}\,
    \bar{f}(\epsilon_{\beta',k'+k''})
    \\
   &+\text{same terms with the replacement}\:
    1, \L \rightarrow N, \R.
    \end{split}
\end{equation}
In order to solve these equations, we have to eliminate
$\rho_{00,k}$ which is most conveniently done by inserting the Fourier
representation of the normalisation condition
\begin{equation}
\label{trace}
\tr_\text{wire}\rho_0(t)
= \rho_{00}(t) + 2\sum_{\alpha} \rho_{\alpha\alpha}(t) = 1 .
\end{equation}

In order to obtain for the current an expression that is
consistent with the restriction to one excess electron, we compute the
expectation values in the current formula \eqref{I(t)ME} with
the reduced density operator \eqref{eq:decomposition} and insert the
Floquet representation \eqref{eq:floquet_operator} of the wire
operators.  Performing an average over the driving period, we obtain
for the dc current the expression \cite{Kaiser2006b}
\begin{equation}
\label{eq:dc-current}
\begin{split}
    \bar I = \frac{2e \Gamma_\L}{\hbar} \Re \sum_{\alpha,k}\Big(
    &
    \sum_{\beta,k'}
    \langle \varphi_{\beta,k'+k} | 1 \rangle
    \langle 1 | \varphi_{\alpha,k} \rangle
    \rho_{\alpha \beta, k'}
    \bar{f}_\L(\epsilon_{\alpha,k})
    \\ - &
    \sum_{k'}
    \langle \varphi_{\alpha,k'+k} | 1 \rangle
    \langle 1 | \varphi_{\alpha,k} \rangle 
    \rho_{00,k'}
    f_\L(\epsilon_{\alpha,k})
    \Big).
\end{split}
\end{equation}
Physically, the second contribution of the current formula
\eqref{eq:dc-current} describes the tunnelling of an electron from the
left lead to the wire and, thus, is proportional to $\rho_{00} f_\L$
which denotes the probability that a lead state is occupied while the
wire is empty.  The first terms corresponds to the reversed process
namely the tunnelling on an electron from site $|1\rangle$ to the left
lead.

The decomposition of the equation of motion \eqref{eomFperp} for the
long-time limit of $\mathcal{F}_{1\perp}$ and the subsequent
computation of the $\bar S$ from Eq.~\eqref{Soperator} proceeds along
the same lines with the only difference that the current operators
$\mathcal{J}_\pm$ yield an inhomogeneity and that the r.h.s.\ of the
trace condition \eqref{trace} is
\begin{equation}
\tr_\text{wire}\mathcal{F}_{1\perp}
= (\mathcal{F}_{1\perp})_{00}
+ 2\sum_{\alpha} (\mathcal{F}_{1\perp})_{\alpha\alpha} = 0.
\end{equation}

\subsection{Spinless electrons}

A particular consequence of strong Coulomb repulsion is the mutual
blocking of different spin channels.  This motivates us to also
compare to the case of spinless electrons which is physically realised
by spin polarisation.  For spinless electrons, we drop in the
Hamiltonian \eqref{eq:H} all spin indices.  Physically, this limit is
realised by a sufficiently strong magnetic field that polarises all
electrons contributing to the transport.
By the same calculation as above, we then obtain for
the current also the expression \eqref{eq:dc-current} but without the
prefactor $2$.  The factor $2$ is also no longer present in the normalisation
condition \eqref{trace} which now reads
\begin{equation}
\label{trace-nospin}
\tr_\text{wire}\rho_0(t)
= \rho_{00}(t) + \sum_\alpha \rho_{\alpha\alpha}(t) = 1 ,
\end{equation}
and accordingly in the equation of motion for $\mathcal{F}_{1\perp}$.

\section{Noise in non-adiabatic electron pumps}
\label{sec:pumping}

In a mesoscopic conductor, a spatial asymmetry together with an ac
driving can induce a pump current, i.e.\ a dc current that flows even
in the absence of any net bias \cite{Kouwenhoven1991a, Pothier1992a,
Switkes1999a}.  More precisely, the central necessary condition for
this effect is the absence of a particular symmetry, namely
generalised parity defined as the invariance under spatial reflection
in combination with a time shift by half a driving period
\cite{Kohler2005a}.
For adiabatically slow driving, the transferred charge is determined
by the area enclosed in parameter space during one cycle of the
periodic time-evolution \cite{Brouwer1998a, Altshuler1999a}.  This
implies that the resulting dc current, apart from non-adiabatic
corrections, is proportional to the driving frequency.  For
time-reversal symmetric conductors, the parameters do not
enclose a finite area and, thus, one finds for small driving
frequencies that the pump current obeys $\bar I\propto \Omega^2$
\cite{Strass2005b}.

In both the absence and the presence of time-reversal symmetry, the
current increases with the driving frequency, which suggests that
pumping is more effective beyond the adiabatic regime.  Non-adiabatic
electron pumping is particularly interesting because at internal
resonances of the central system the pump current can assume rather
large values \cite{Stafford1996a, Brune1997a, vanderWiel2003a}.
In the absence of electron-electron interactions, the pump current, in
addition, exhibits resonance peaks with a remarkably low current noise
\cite{Strass2005b} and the question arises whether this favourable
property persists once Coulomb repulsion becomes relevant.
We consider the setup sketched in Fig.~\ref{fig:levels} and described
by the Hamiltonian \eqref{eq:H} and compare the results for
non-interacting electrons with those for strong Coulomb repulsion.
Thereby we focus on the parameter set for which recently a low Fano
factor has been predicted \cite{Strass2005b}: It is characterised by a
large internal bias, intermediately strong wire-lead coupling, and
resonant driving.  Moreover at the prime resonance, the driving
amplitude has to fulfil the condition $J_1(A/\hbar\Omega) =
\sqrt{5/3}\Gamma_\mathrm{L,R}$, where $J_1$ denotes the first-order
Bessel function of the first kind.

\begin{SCfigure}[4][tb]
\includegraphics{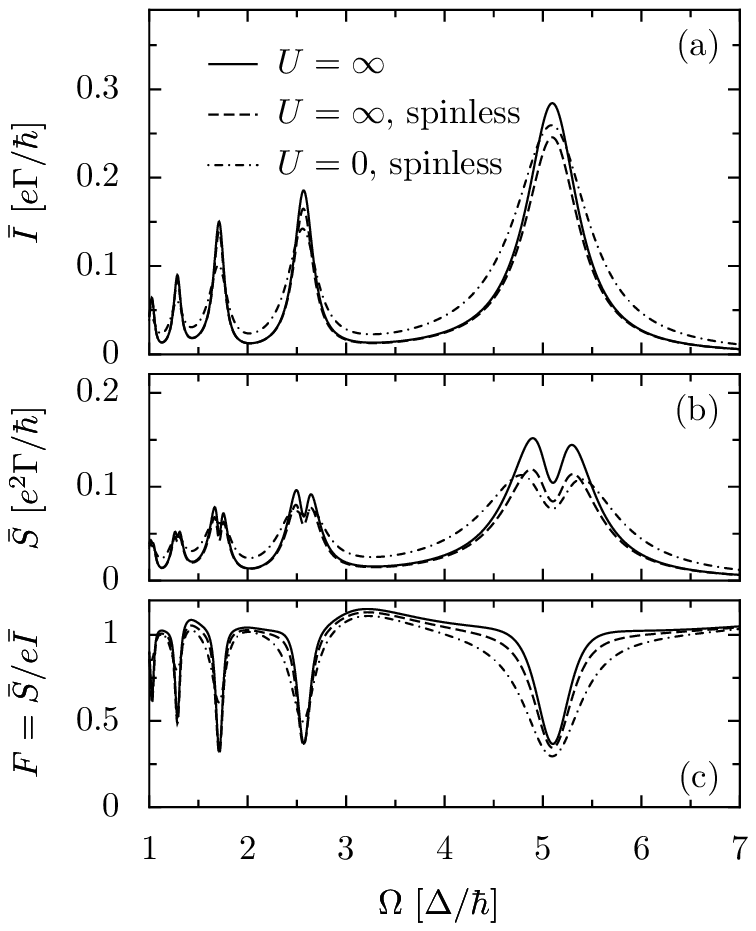}%
\caption{DC current (a), zero-frequency noise (b), and Fano factor
(c) for the non-adiabatically driven electron pump sketched in
Fig.~\ref{fig:levels} as a function of the driving frequency.
The dot levels possess the energies $\epsilon_{1,2}=\pm 2.5\Delta$ and
are coupled to the leads with the strength $\Gamma_{\mathrm{L,R}} 
=0.3\Delta$.  The driving amplitude is $A=3.7\Delta$ and the
temperature $k_\mathrm{B}T=0.005\Delta$. The peak at $\Omega\approx
\Delta/\hbar$ corresponds to the first-order resonance, while the
peaks at lower frequencies are higher-order resonances.
\par\quad\par\quad\par}
\label{fig:ISF}
\end{SCfigure}

Figure~\ref{fig:ISF}a shows the current obtained for non-interacting
electrons and for strong Coulomb repulsion with and without
considering the spin degree of freedom.  The dc current exhibits
characteristic peaks whenever the $k$th-order resonance condition
$\Omega = [(\epsilon_1-\epsilon_2)^2 +\Delta^2]^{1/2}/k\hbar$ is
fulfilled.  Interestingly enough, the magnitude of the current is
practically the same for all three cases: The interaction yields at
most differences of the order 10\%.  We also find that for $U=\infty$,
the resonance peaks are slightly sharper, in particular if one
takes the electron spin into account.  This behaviour has also been
found for photo-assisted transport in molecular wires, where it is
even significantly more pronounced \cite{Kaiser2006b}.
The zero-frequency noise shown in Fig.~\ref{fig:ISF}b possesses
a more involved double-peak structure with a local minimum at the
centre.  This results in a rather pronounced minimum of the Fano
factor (Fig.~\ref{fig:ISF}c) which assumes values as low as
$F\approx0.25$, which means sub-Poissonian current noise.

Our observations lead us to the conclusion that interactions are not
an essential obstacle for tuning the double-dot electron pump into a
low-noise regime as suggested in Ref.~\cite{Strass2005b}.  The reason
for this is that for the pump configuration skechted in
Fig.~\ref{fig:levels}, one energy level lies below the Fermi energy
while the other lies well above.  Consequently in equilibrium for a
sufficiently small wire-lead coupling, the left site is occupied while
the right site is empty, whatever the interaction strength.  Thus, the
double dot is populated with only one electron so that interactions
become irrelevant.  Unless the driving amplitude is huge, this
occupation is altered only slightly.  Consequently interactions do not
modify the current significantly.
We emphasise that for strong dot-lead coupling $\Gamma$ and finite
interaction $U$, these arguments no longer hold true.

In the experiment of Ref.~\cite{vanderWiel1999a}, a typical inter-dot
coupling is $\Delta = 50\mu\mathrm{eV}$.  Then, an internal bias
$\epsilon_0=5\Delta$ corresponds to the resonance frequency
$\Omega=5\Delta/\hbar \approx 2\pi\times 60\mathrm{GHz}$.  In both
interacting and non-interacting electrons, a wire-lead coupling
$\Gamma=0.1\Delta$ results in an optimised pump current of the order
$200\mathrm{pA}$ with a Fano factor $F\approx0.25$.

\section{Conclusions}

Within the present work, we were particularly interested in the
question whether Coulomb repulsion would deteriorate the desirable
noise properties of the pump current found for non-interacting
electrons.  As a central result, we found that very strong
interactions do not alter the picture, but merely lead to slight
modifications of the current and the zero-frequency noise, which are
of the order of~10\%.  Therefore the essential conclusion of
Ref.~\cite{Strass2005b} holds also true in the strongly interacting
case: Low-noise operation of a non-adiabatic electron pump requires a
large internal bias in combination with a strong inter-dot coupling
and resonant driving.  The consequence is that a properly tuned
driving amplitude provides a relatively large pump current with
clearly sub-Poissonian noise.  Furthermore, we found that the
influence of the electron spin is also not significant.

For non-interacting electrons, the transport problem can be treated
within a Floquet scattering formalism, which relates the dc current to
the transmission probabilities of elastic and inelastic transmission
channels.  The latter account for the possibility of photon emission
and absorption by a scattered electron.  The zero-frequency noise, by
contrast, is not only determined by probabilities, but also depends on
the scattering phases.

As soon as electron-electron interactions play a significant role, a
scattering approach becomes of limited use.  In the case of weak
wire-lead coupling, the transport problem can then be treated within a
master equation approach that is based on second order perturbation
theory in the coupling.  While previous master equation studies of
driven transport are restricted to the computation of the dc current,
we developed a method to also compute the noise properties.
Our method is based on a time-dependent generalised master equation
for the full counting statistics.  Concerning its solution, the
time-dependence of the Liouville operator represents a particular
challenge, because it renders the asymptotic state of the system
time-dependent as well.  We coped with this difficulty by decomposing
the master equation into a proper Floquet basis.

Concerning the noise of non-adiabatic electron pumps, still
a couple of intriguing questions remain. Thus far, we revealed that
the two limiting cases of vanishing and strong interaction allow for
low-noise pumping, but the transition between these limits may
nevertheless bear surprises.  Thus studying current noise for
a finite interaction strength is highly desirable.
Moreover, albeit for ac driven conductors experimentally beyond the
present state of the art, cumulants of higher order and eventually the
full counting statistics can be of interest.
In that spirit, our study can only be a first step towards a more
complete characterisation of the electron transport in the presence of
Coulomb interaction and time-dependent fields.

\begin{acknowledgement}
It is a pleasure to thank Peter H\"anggi for interesting discussions.
This work has been supported by DFG through SPP 1243 ``Quantum transport
on the molecular scale''.
Financial support of the German Excellence Initiative
via the ``Nanosystems Initiative Munich (NIM)'' and of the
Elite Network of Bavaria via the International 
Doctorate Program ``NanoBioTechnology'' is gratefully acknowledged.
\end{acknowledgement}

\bibliographystyle{prsty}

\end{document}